## 1. Introduction

Digital topology studies geometric and topological properties of digital images arising in many areas of science including fluid dynamics, geoscience, neuroscience and medical imaging. In particular, integrating geometric and topological constraints into discretization schemes in order to generate geometrically and topologically correct digital models of anatomical structures is critical for many clinical and research applications (e.g. [14-15]). Although many clinical and research applications require accurate segmentations, only a few techniques have been proposed to achieve accurate and geometrically and topologically correct segmentations.

Discretization schemes preserving geometry and topology of the object of interest were introduced in [8]. These schemes enable to obtain more detailed geometric and topological information about the specific regions of the object. The shape and the size of individual elements can be arbitrary within the framework of the proposed scheme that allows representing objects with fine anatomical details. Another feature is that the digital model of a 2-dimensional continuous object is necessarily a digital 2-surface preserving the topology of the object. The approach studied in this paper is based on locally centered lump (LCL) discretization of n-dimensional objects.

Over the last decade, there have been proposed numerous new models devoted to applying geometric and topological tools to digital image analysis.

Usually, a digital object is equipped with a graph structure based on the local adjacency relations of digital points [5]. In papers [6-7], a digital n-surface was defined as a simple undirected graph and basic properties of n-surfaces were studied. Paper [6] analyzes a local structure of the digital space $Z^n$. It is shown that $Z^n$ is an n-surface for all n>0. In paper [7], it is proven that if A and B are n-surfaces and A⊆B, then A=B.

X. Daragon et al. [3-4] studied partially ordered sets in connection with the notion of n-surfaces. In particular, it was proved that (in the framework of simplicial complexes) any n-surface is an n-pseudomanifold, and that any n-dimensional combinatorial manifold is an n-surface. An interesting method using cubical images with direct adjacency for determining such topological invariants as genus and the Betti numbers was designed and studied by L. Chen et al. [2].

In this paper, we investigate connection between closed surfaces and their digital models. Section 2 studies LCL collections of 2-dimensional cells. In section 3, we use digital 2-dimensional surfaces that were studied in [6-7]. We introduce the notion of a compressed digital 2-dimensional surface and show that a digital 2-dimensional surface can be converted to a compressed form by transformations retaining the connectedness and the dimension of the given digital 2-surface. In section 4, the digital model G(M) of a closed surface M is defined. The digital model G(M) is the intersection graph of an LCL cover of M. It is shown that any two digital models of M are homotopy equivalent digital 2-manifolds. In sections 5-6, we establish a link between closed surfaces and compressed digital 2-surfaces, and classification of closed surfaces by digital tools.

## 2. LCL collections of 1- and 2-cells

In this section, we use intrinsic topology of an object, without reference to an embedding space. We say that a set D is a 2-cell if it is homeomorphic to a closed unit square, a set D is a 1-cell if it is homeomorphic to a closed unit segment, a set C is a circle or a 1-sphere if it is homeomorphic to a unit circle. We denote the interior and the boundary of an n-cell D, by IntD and ∂D respectively. Note that D=IntD∪∂D. The boundary of a 1-cell is two endpoints, the boundary of a 2-cell is a circle. The 0-cell D is a single point for which ∂D=∅. Facts about n-cells that we will need in this paper are stated below.

**Facts 2.1**
- If C is a circle and D is a 1-cell contained in C then C-IntD is a 1-cell.
- Let $C_1$ and $C_2$ be 1-cells such that $C_1 \cap C_2 = \partial C_1 \cap \partial C_2 = v$ is an endpoint of $C_1$ and $C_2$. Then $C_1 \cup C_2 = E$ is a 1-cell.
- Let $D_1$ and $D_2$ be 2-cells such that $D_1 \cap D_2 = \partial D_1 \cap \partial D_2 = C$ is a 1-cell. Then $D_1 \cup D_2 = B$ is a 2-cell.

**Definition 2.1**
Let $W=\{X_1,\ldots X_s\}$ be a collection of sets. W is called a locally centered collection (LC collection) if for all $m,k \in Q \subseteq \{1,\ldots s\}$, from condition $X_k \cap X_m \neq \emptyset$, $m \neq k$, it follows that $\cap\{X_k : k \in Q\} \neq \emptyset$.

In topology, this property is often called Helly's property. In application to digital topology, collections of sets with similar properties were studied in a number of works.

**Definition 2.2**
1. Let $W=\{C_1,C_2,\ldots\}$ be a collection of 1-cells. W is called a locally lump collection (LL collection) if:
   a) From condition $C_k \cap C_m \neq \emptyset$, $m \neq k$, it follows that $C_k \cap C_m = \partial C_k \cap \partial C_m = v$ is a point.
   b) The intersection of any three distinct 1-cells is empty.
2. W is called a locally centered lump collection (LCL collection) if W is a locally centered collection and a locally lump collection at the same time (fig. 1).

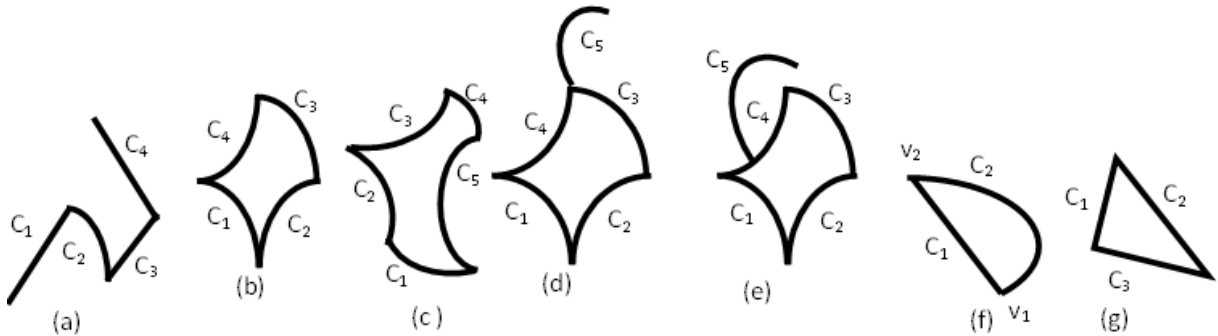

Figure 1. Collections of 1-cells. (a)-(c) are LCL collections. (d)-(f) are not LCL collections.

Collections of 1-cells are shown in fig. 1. Collections (a)-(c) are LCL collections. (d) is not an LCL collection because $C_3 \cap C_4 \cap C_5 \neq \emptyset$. (e) is not an LL collection because $C_4 \cap C_5 \neq \partial C_4 \cap \partial C_5$. (f) is not an LL collection because $C_1 \cap C_2 = \{v_1, v_2\}$. (g) is not an LCL collection because $C_1 \cap C_2 \cap C_3 = \emptyset$. Clearly, any subcollection of an LCL collection of 1-cells is an LCL collection.

**Definition 2.3**
1. Let $W=\{D_1,D_2,\ldots\}$ be a collection of 2-cells. W is called a locally lump collection (LL collection) if:
a)  From condition $D_k \cap D_m \neq \emptyset$, $m \neq k$, it follows that $D_k \cap D_m = \partial D_k \cap \partial D_m = C_{km}$ is a 1-cell.
b)  From condition $D_k \cap D_m \cap D_p \neq \emptyset$, $m \neq k$, $m \neq p$, $p \neq k$, it follows that
$D_k \cap D_m \cap D_p = \partial D_k \cap \partial D_m \cap \partial D_p = C_{km} \cap C_{kp} = \partial C_{km} \cap \partial C_{kp} = v$ is a point.
c)  The intersection of any four distinct 2-cells is empty.
2. W is called a locally centered lump collection (LCL collection) if W is a locally centered collection and a locally lump collection at the same time (fig. 2).

LCL collections of 1- and 2-cells were defined and studied in [8]. Collections of 2-cells are depicted in fig. 2. Collection (a) is not an LCL collection, because $D_1 \cap D_3$ and $D_2 \cap D_4$ are not 1-cells. Collection (b) is not an LCL collection because $D_1 \cap D_2 \neq \emptyset$, $D_1 \cap D_3 \neq \emptyset$, $D_2 \cap D_3 \neq \emptyset$ but $D_1 \cap D_2 \cap D_3 = \emptyset$. Collections (c)-(e) are LCL collections of 2-cells. Obviously, any subcollection of an LCL collection of 2-cells is an LCL collection. In paper [7], a locally centered collection is called continuous and it is shown that for a given object, the intersection graphs of all continuous, regular and contractible covers are homotopy equivalent to each other. In papers [1, 13], a normal set W of convex nongenerate polygons (intersection of any two of them is an edge, a vertex, or empty) is called strongly normal (SN) if for all $P, P_1,\ldots P_n$ $(n>0) \in W$, if each $P_i$ intersects P and $I = P_1 \cap \ldots \cap P_n$ is nonempty, then I intersects P. Several papers, e.g. [11-12] extended basic results about strong normality to collections of polyhedra in $R^n$. There are obvious differences between SN collections of polygons and LCL collections. For example, elements of an

SN collection are convex sets whereas any 2-cell in an LCL collection can be of an arbitrary shape and size. Let us remind the definition of isomorphic sets. A collection W={$A_0,A_1,...$} of sets is isomorphic to a collection V={$B_0,B_1,...$} of sets, if there exists one-one onto correspondence f: W→V such that $A_i \cap ... A_k \cap ... A_p \neq \emptyset$ if and only if $f(A_i) \cap ... f(A_k) \cap ... f(A_p) \neq \emptyset$. The following assertion was proven in [8].

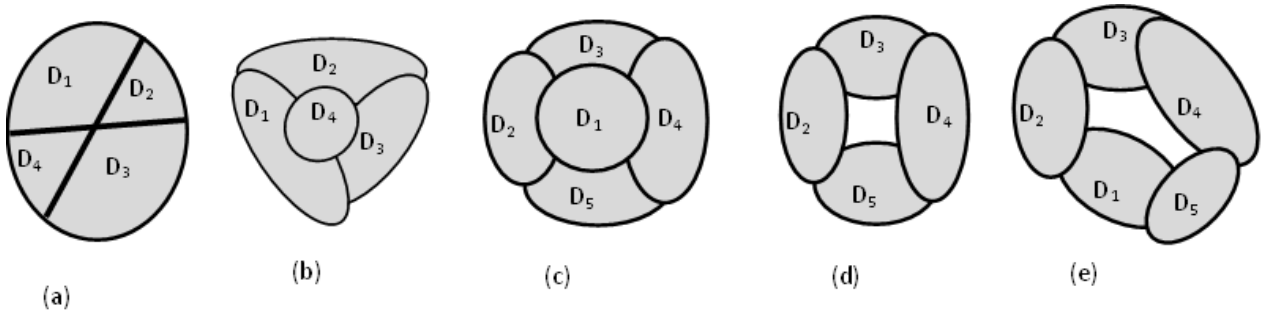

Figure. 2. (a)-(b) are not LCL collections. (c)-(e) are LCL collections of 2-cells.

**Proposition 2.1**
Let W={$D_0,D_1,...$} be an LCL collection of 2-cells, U={$D_1,D_2,...D_s$} be a collection of all 2-cells intersecting $D_0$, and V={$C_1,C_2,...C_s$} be the collection of 1-cells $C_i = D_0 \cap D_i \neq \emptyset$, i=1,...s. Then:
- V is an LCL collection.
- Collections U and V are isomorphic, and f: U→V, $f(D_i)=C_i$, i=1,...s, is an isomorphism.

### 3. Contractible graphs and contractible transformations. Digital manifolds

Traditionally, a digital space G is a simple undirected graph G=(V,W), where V={$v_1,v_2,...v_n,...$} is a finite or countable set of points, and W = {$(v_p v_q),....$}⊆V×V is a set of edges. Such notions as the connectedness, the adjacency, the dimensionality and the distance on a graph G are completely defined by sets V and W. Further on, if we consider a graph together with the natural topology on it, we will use the phrase 'digital space". We use the notations $v_p \in G$ and $(v_p v_q) \in G$ if $v_p \in V$ and $(v_p v_q) \in W$ respectively if no confusion can result. |G| denotes the number of points in G.

Since in this paper we use only subgraphs induced by a set of points, we use the word subgraph for an induced subgraph. We write H⊆G. Let G be a graph and H⊆G. G-H will denote a subgraph of G obtained from G by deleting all points belonging to H. For two graphs G=(X,U) and H=(Y,W) with disjoint point sets X and Y, their join G⊕H is the graph that contains G, H and edges joining every point in G with every point in H. Points $v_p$ and $v_q$ are called adjacent if $(v_p v_q) \in W$. The subgraph O(v)⊆G containing all points adjacent to v (without v) is called the rim or the neighborhood of point v in G, the subgraph U(v)=v⊕O(v) is called the ball of v. Graphs can be transformed from one into another in a variety of ways. Contractible transformations of graphs seem to play the same role in this approach as a homotopy in algebraic topology [10].

A graph G is called contractible (fig. 3) if it can be converted to the trivial graph by sequential deleting simple points. A point v of a graph G is said to be simple if its rim O(v) is a contractible graph.

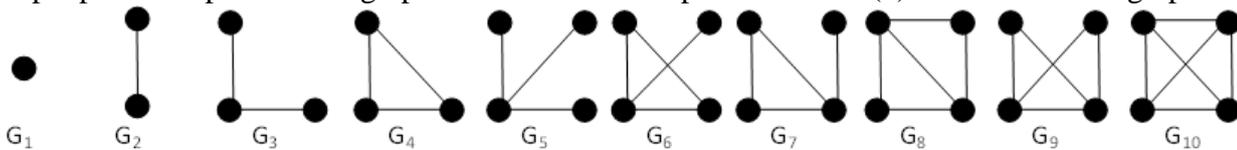

Figure 3. Contractible graphs with the number of points n<5.

An edge (vu) of a graph G is said to be simple if the joint rim O(vu)=O(v)∩O(u) is a contractible graph. In [10], it was shown that if (vu) is a simple edge of a contractible graph G, then G-(vu) is a contractible graph. Thus, a contractible graph can be converted to a point by sequential deleting simple points and edges. In fig.3, $G_{10}$ can be converted to $G_9$ or $G_8$ by deleting a simple edge, and can be converted to $G_4$ by deleting a simple point. $G_9$ can be converted to $G_7$ or $G_6$ by deleting a simple edge. $G_6$ can be converted to $G_5$ by deleting a simple edge. $G_7$ can be converted to $G_4$ by deleting a simple point. $G_5$ can be converted to

$G_3$ by deleting a simple point. $G_3$ can be converted to $G_2$ by deleting a simple point. $G_2$ can be converted to $G_1$ by deleting a simple point.

Deletions and attachments of simple points and edges are called contractible transformations. Graphs G and H are called homotopy equivalent or homotopic if one of them can be converted to the other one by a sequence of contractible transformations.

Homotopy is an equivalence relation among graphs. Contractible transformations retain the Euler characteristic and homology groups of a graph [10].

Properties of graphs that we will need in this paper were studied in [7, 10].

**Proposition A**
- Let G be a graph and v be a point ($v \notin G$). Then $v \oplus G$ is a contractible graph. If K is a clique then $K \oplus G$ is a contractible graph.
- Let G be a contractible graph, and S(a,b) be a disconnected graph with just two points a and b. Then $S(a,b) \oplus G$ is a contractible graph.
- Let G be a contractible graph with the cardinality $|G|>1$. Then it has at least two simple points.
- Let H be a contractible subgraph of a contractible graph G. Then G can be transformed into H sequential deleting simple points.

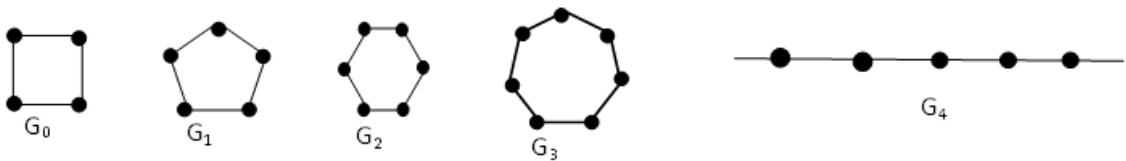

Figure. 4. $G_0$ - $G_3$ are digital closed 1-surfaces (digital circles). $G_4$ is the digital line. $G_5$-$G_8$ are digital closed 2-surfaces. $G_9$ is a portion of the digital plane.

There is an abundant literature devoted to the study of different approaches to digital lines surfaces and spaces used by researchers, just mention some of them [1-5, 11-13]. Digital n-manifolds were defined and investigated in [6-9]. Consider now digital 1- and 2-manifolds.

**Definition 3.1**
- A digital 0-dimensional sphere is a disconnected graph $S^0(a,b)$ with just two points a and b.
- A connected graph M is called a digital 1-manifold, if the rim O(v) of each point v of M is a digital 0-sphere (fig. 4).
- Let M be a digital 1-manifold and x and y be adjacent points of M. We say that {x,y} is a simple pair if the space $(U(x) \cup U(y))-\{x,y\}$ is a digital 0-sphere {fig. 5(a)).
- Let M be a digital 1-manifold, and {x,y} be a simple pair lying in M. The contraction of points x and y in M is the replacement of x and y with a point z such that z is adjacent to the points to which points x and y were adjacent (fig. 5(b-c).

The minimal number of points of a digital 1-manifold is four. It is a minimal digital 1-sphere. Digital 1-manifolds $G_0$-$G_4$ are depicted if fig. 4. $G_0$-$G_3$ are digital 1-manifolds with a finite number of points. $G_0$ is a minimal digital 1-sphere. A digital line $G_4$ has an infinite number of points. In fig. 5(a), {x,y} is a simple pair because $(U(x) \cup U(y))-\{x,y\}=S^0(u,v)$ is a pair of two non-adjacent points.

Note that in graph theory, the contraction of points x and y in a graph G is the replacement of x and y with a point z such that z is adjacent to the points to which points x and y were adjacent.

**Definition 3.2.**
- The join $S^1_{min}= S^0(u_1,v_1) \oplus S^0(u_2,v_2)$ of two copies of a zero-dimensional sphere is called a minimal digital 1-sphere (see $G_0$ in fig. 4)
- A digital 1-manifold M is called a digital 1-sphere if M can be converted to $S^1_{min}$ by a sequential

contraction of simple pairs (fig. 5(a-c)).

In fig. 5(a-c), M is a digital 1-sphere with six points, {x,y} is a simple pair of points, and N=(M∪z)-{x,y} is a digital 1-sphere with five points.

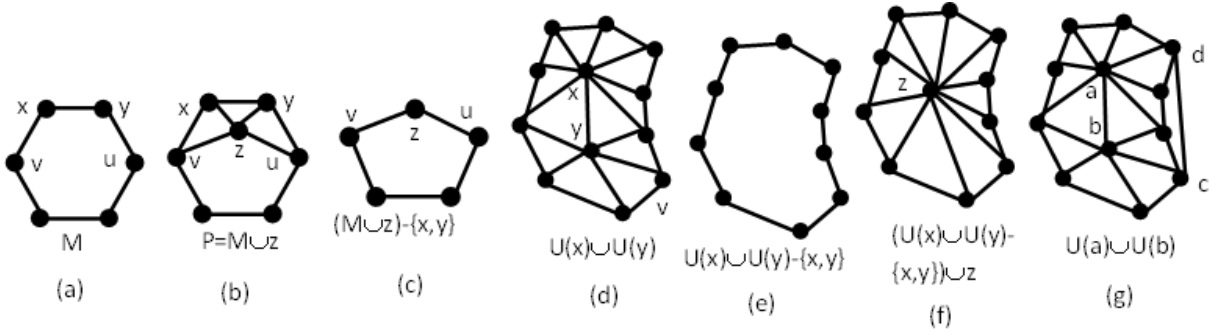

Figure 5. (a) M is a digital 1-sphere with six points. (b) {x,y} is a topological pair of points. (c) N=(M∪z)-{x,y} is a digital 1-sphere with five points. (d) {x,y} is a simple pair. U(x)∪U(y) is a contractible space. (e) U(x)∪U(y)-{x,y} is a digital 1-sphere. (f) The rim of z is a digital 1-sphere. (g) {a,b} is not a topological pair of points. U(a)∪U(b)-{a,b} is not a digital 1-sphere.

The following assertion was proven in [9].

**Proposition 3.1**
- Let {x,y} be a simple pair lying in a digital 1-manifold M. Then the space N=(M∪z)-{x,y} obtained by the contraction of {x,y} is a digital 1-manifold homotopy equivalent to M.
- A digital 1-manifold M with a finite number of points is a digital 1-sphere.
- A compressed digital 1-manifold is a minimal 1-sphere.

**Definition 3.3**
- A connected graph M is called a digital 2-manifold if for each point v of M, the rim O(v) is a digital 1-sphere [6], (see fig. 6).
- Let M be a digital 2-manifold and x and y be adjacent points of M. We say that {x,y} is a simple pair if the space (U(x)∪U(y))-{x,y} is a digital 1-sphere (fig. 5(d)).
- Let M be a digital 2-manifold, and {x,y} be a simple pair lying in M. The contraction of points x and y in M is the replacement of x and y with a point z such that z is adjacent to the points to which points x and y were adjacent (fig, 5(d-f)).

**Definition 3.4.**
- The join $S^2_{min}= S^0(u_1,v_1)\oplus S^0(u_2,v_2) \oplus S^0(u_3,v_3)$ of three copies of a 0-dimensional sphere is called a minimal digital 2-sphere ($G_1$ in fig. 6).
- A digital 2-manifold M is called a digital 2-sphere if M can be converted to $S^2_{min}$ by sequential contracting simple pairs (spaces $G_1$-$G_3$ in fig, 6).

A pair {x,y} is simple in digital 2-manifolds shown in fig. 5(d-f) and in fig. 6 in $G_2$. A pair {a,b} depicted in fig. 5 (g) is not a simple pair. The following assertion was proven in [9].

**Proposition 3.2**
Let {x,y} be a simple pair lying in a digital 2-manifold M. Then the space N=(M∪z)-{x,y} obtained by contraction of {x,y} is a digital 2-manifold homotopy equivalent to M (see $G_1$ and $G_2$ in fig. 6).

It follows from propositions 3.1 and 3.2 that the number of points in a digital n-manifold M can be reduced by using contractible transformations, which retain the topology of M.

**Definition 3.5**
A digital n-manifold M, n=1,2, is called compressed if it has no simple pairs.

Obviously, a compressed digital 1-manifold is a minimal 1-sphere, and a compressed digital 2-sphere is a minimal 2-sphere (see $G_1$ in fig 4 and 6) . For any digital 2-manifold M there exists a compressed digital 2-manifold C(M) with the minimal number of points. A digital 2-manifold M can be converted to a compressed form by sequential contracting simple pairs. In paper [9], it was shown that if N is a compressed digital n-manifols, n>0, and a digital n-manifold M is homotopy equivalent to N, then M can be converted to N by sequential contracting simple pairs. The following proposition was proven in [9] (proposition 5.2).

**Proposition 3.3**
- Let M be a digital 2-manifold with a finite number of points, and G be a compressed digital 2-manifold homotopy equivalent to M. Then M can be converted to G by sequential contracting simple pairs.
- Let M be a compressed digital 2-manifold, and points v and u be adjacent in M. Then there is a digital minimal 1-sphere S consisting of four points, lying in M, and containing points v and u.

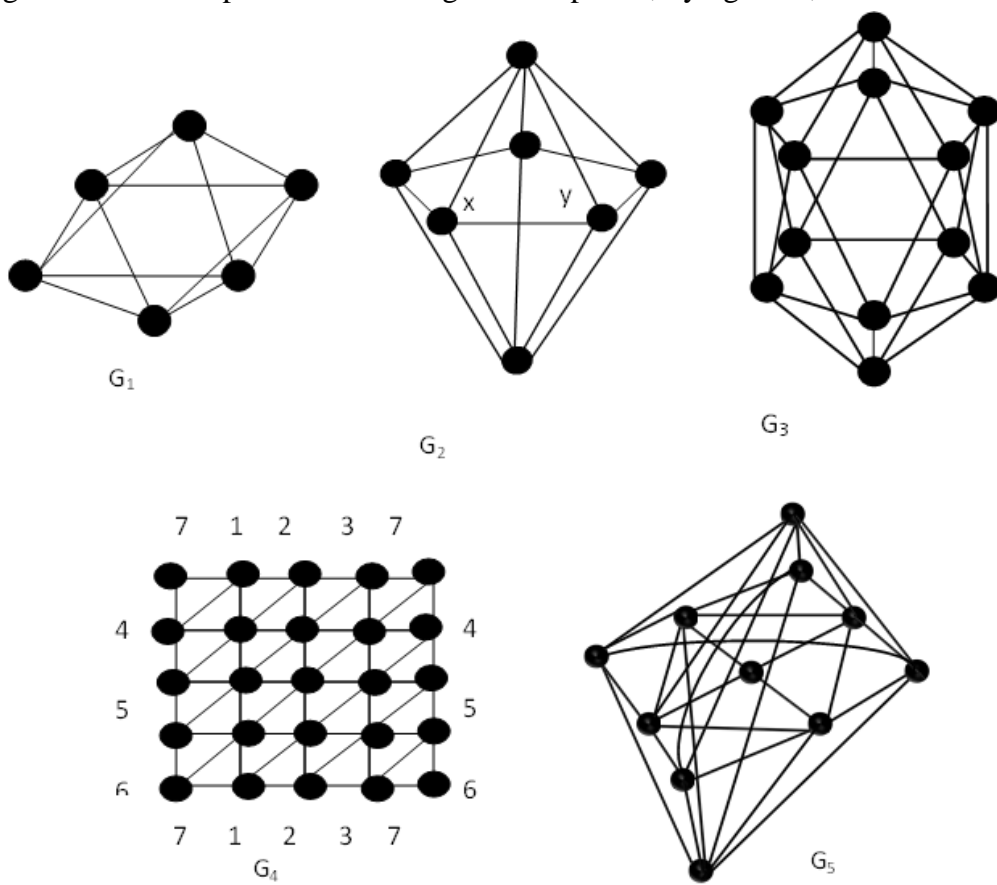

Figure. 6. $G_1 - G_5$ are digital 2-manifolds. $G_1$-$G_3$ are digital 2-spheres. $G_4$ is a digital 2-torus. $G_5$ is a digital 2-projective plane.

In fig.6, the compressed 2-sphere $G_1$ contains six points, the compressed 2-torus $G_4$ contains sixteen points, the compressed two-dimensional projective plane $G_5$ contains eleven points. Suppose that W(M)={M, N, P,…} is a set of digital 2-manifolds homotopy equivalent to M, and C(M) is a compressed 2-manifold obtained by sequential contracting simple pairs lying in M. According to proposition 3.3, any 2-manifold Q belonging to W(M) can be transformed to C(M) by a sequential contraction of simple pairs. Therefore, C(M) can represent the class W(M) of all manifolds contained in W(M).

**Definition 3.6**

Let M be a digital 2-manifold with a finite number of points, and let C(M) be a compression of M, i.e., a digital compressed 2-manifold homotopy equivalent to M. The digital weight of M, denoted by dw(M) is the number |C(M)| of points of M.

**4. Intersection graphs of LCL coves of curves and closed surfaces and digital models of closed surfaces**

**I**n this section, we investigate a connection between LCL covers of closed surfaces and intersection graphs of the covers. We are going to show that an LCL cover of a closed surface incorporates geometric and topological features directly into the digital model of this surface.

Remind the definition of the intersection graph of a family of sets. Let $W=\{D_1,D_2,\ldots D_n,\ldots\}$ be a finite or countable family of sets. Then the graph $G(W)$ with points $\{x_1,x_2,\ldots x_n,\ldots\}$ is called the intersection graph of W, if points $x_k$ and $x_i$ are adjacent whenever $D_k \cap D_i \neq \emptyset$. In other word, f: $G(W) \to W$ such that $f(x_i)=D_i$ is an isomorphism.

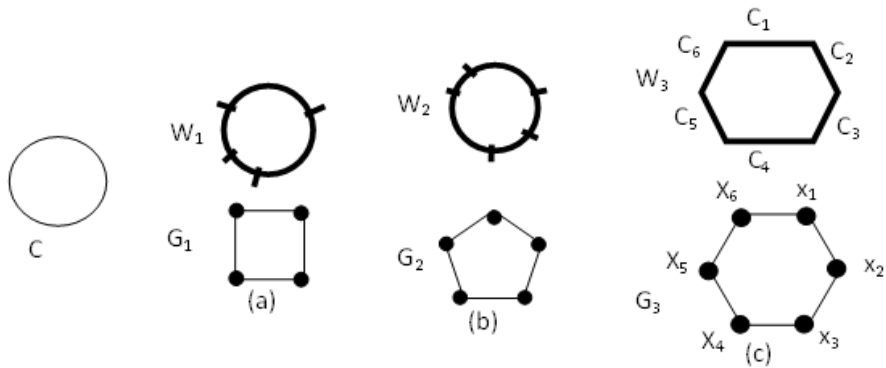

Figure 7. $W_1$, $W_2$, $W_3$ are LCL covers of a circle C, $G_1$, $G_2$, $G_3$ are the corresponding digital models of C.

**Definition 4.1**
Let M be a circle, $W(M)=\{C_0,C_1,C_2,\ldots,C_n\}$ be an LCL cover of $M=C_0\cup\ldots C_n$ by 1-cells, and $G(W(M))=\{x_0,x_1,x_2,\ldots,x_n\}$ be the intersection graph of W(M). Then G(W(M)) is called the digital model of M in regard to W(M).

In fig 7, $W_1$, $W_2$, $W_3$ are LCL covers of a circle C, $G_1$, $G_2$, $G_3$ are the corresponding digital models of C. Obviously, $G_1$, $G_2$, $G_3$ are digital 1-spheres. The following theorem was proven in [8].

**Theorem 4.1**
Let M be a circle, and $W(M)=\{C_0,C_1,\ldots,C_n\}$ be an LCL cover of M by 1-cells. Then the digital model G(W(M)) of M is a digital 1-sphere.

**Remark 1**
Suppose that D is a 2-cell in Euclidean space $E^2$. Divide $E^2$ into a set $V=\{u_1,u_2,\ldots\}$ of closed squares with edge length L, and vertex coordinates equal to nL. V is a cover of $E^2$. Pick out the family $V(D)=\{u_1,\ldots u_s\}$ of squares intersecting D. V(D) is a cover of D. Then build the intersection graph G(V(D)) of V(D). Since D is a topological closed 2-disk then for small enough L, the union $E=u_1\cup\ldots u_s$ is a 2-cell, and the intersection graph $G(V(D))=\{x_1,\ldots x_s\}$ of V(D) is contractible (see fig. 8(a)-(c)). For the same reason, if D is a 1-cell then $G(V(D))=\{x_1,\ldots x_s\}$ of V(D) is a contractible graph.

**Theorem 4.2**
Let M be a closed surface, and collections $W_1(M)=\{D_1,D_2,\ldots D_m\}$ and $W_2(M)=\{E_1,E_2,\ldots E_n\}$ be LCL covers of M by 2-cells. Then the intersection graphs $G(W_1(M))$ and $G(W_2(M))$ are homotopy equivalent to each other.
**Proof.**

A closed surface M can be formed in the plane from a fundamental polygon P by gluing corresponding sides of the boundary together. Consider P as a unit square shown in fig. 8(d). For definiteness, suppose that M is the connected sum of 3 copies of the projective plane $RP_1\#RP_2\#RP_3$, i.e., M can be represented by P with the string AABBCC as it is shown in fig 8(d).
(1) With no loss of generality, assume that a collection $W_1(P)=\{D_1,D_2,\ldots D_s\}$ of 2-cells depicted in fig.

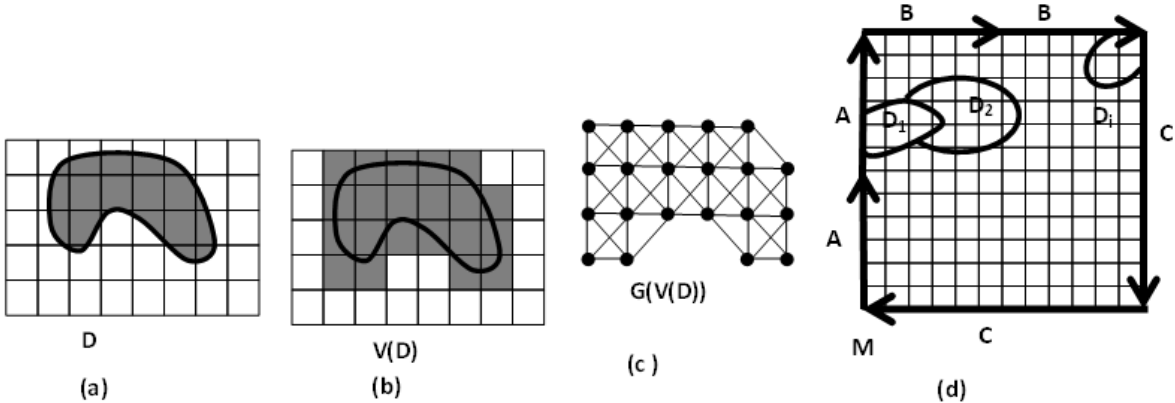

Figure. 8. D is a 2-cell lying in $E^2$. $V(D)=\{u_1,\ldots u_{21}\}$ is a collection of squares intersecting D. $G(V(D))=\{x_1,\ldots x_{21}\}$ is the intersection graph of $V(D)$. $M= RP_1\#RP_2\#RP_3$.

8(d) is an LCL cover of P such that the intersection $IntD_i \cap \partial P=\emptyset$, i=1,…s.
Divide P into a set $V=\{u_1,u_2,\ldots u_s\}$ of closed squares with edge length L=1/n, and vertex coordinates equal to (k/n)L, k=0,1,…n, as it is shown in fig. 8(d). Obviously, V is a cover of P by closed squares, and $G(V(P)))=\{y_{10},\ldots y_{k0}\}$ is the intersection graph of V(P).
Pick out the families $V(D_i)=\{u_{1i},\ldots u_{ki}\}$, i=1,…s, of squares intersecting $D_i$. Then there is a positive $L_1$ such that if $L<L_1$ then the unions $E_i=u_{1i}\cup\ldots u_{ki}$ are 2-cells, and the intersection graphs $G(V(D_i))$ are contractible. For the same reason, if $D_k\cap D_i\neq\emptyset$, (or $D_k\cap D_i\cap D_p\neq\emptyset$), i.e., a 1-cell, and $V(D_k\cap D_i)$ is the family of squares intersecting $D_k\cap D_i$ then the intersection graph $G(V(D_k\cap D_i))$ is contractible.
Glue a simple point $x_i$ to G(V(P)) in such a way that $O(x_i)=G(V(D_i))$, i=1,…s. Then the obtained graph $G_1=G(V(P))\cup\{x_1\cup\ldots x_s\}$ is homotopy equivalent to G(V(P)) by construction. Suppose that $D_k\cap D_i\neq\emptyset$. Then in $G_1$, the joint rim $O(x_kx_i)$ is a contractible graph $G(V(D_k\cap D_i))$ by construction. This means that we can glue a simple edge $(x_ix_k)$ to $G_1$, and the obtained graph $G_2=G_1\cup(x_ix_k)$ is homotopy equivalent to $G_1$. Suppose that $D_n\cap D_m\neq\emptyset$. It is easy to check that the joint rim $O(x_nx_m)$ is a contractible graph in $G_2$. This means that we can glue a simple edge $(x_nx_m)$ to $G_2$, and the obtained graph $G_3=G_2\cup(x_nx_m)$ is homotopy equivalent to $G_2$. In such a way we can glue all simple edges corresponding to all $D_k\cap D_i\neq\emptyset$. Evidently, the obtained graph H is homotopy equivalent to G(V(P)). Note that G(V(P)) is a subgraph of H.
Consider a point $y_{i0}$ belonging to H. Since $y_{i0}$ is adjacent to some point $x_k$ then the rim $O(y_{i0})$ in H is a cone $x_k\oplus A_k$ by construction, i.e., a contractible graph. Therefore, $y_{i0}$ is a simple point, and can be deleted from H. For the same reason, we can delete all points $y_{i0}$, i=1,…k. The obtained graph $H-\{y_{10},\ldots y_{k0}\}= G(W_1(P))=\{x_1,\ldots x_s\}$ is homotopy equivalent to G(V(P)).
(2) Let $W_2(P)=\{F_1,\ldots F_t\}$ be an LCL cover of P by 2-cells, $G(W_2(P))=\{z_1,\ldots z_t\}$ be the intersection graph of $W_2(P)$. For the same reason as above, there is a positive $L_2$ such that if $L<L_2$ then $G(W_2(P))=\{z_1,\ldots z_t\}$ is homotopy equivalent to G(V(P)).
(3) Take $L<\min(L_1,L_2)$. Then $G(W_1(P))=\{x_1,\ldots x_s\}$ is homotopy equivalent to G(V(P)), and $G(W_2(P))=\{z_1,\ldots z_t\}$ is homotopy equivalent to G(V(P)). Therefore, $G(W_1(P))=\{x_1,\ldots x_s\}$ is homotopy equivalent to $G(W_2(P))=\{z_1,\ldots z_t\}$. The proof is complete. □

Fig. 9 shows LCL covers $W_1$-$W_4$ of a continuous sphere S, and the intersection graphs $G(W_1)$-$G(W_4)$ of these covers. Evidently, $G(W_1)$-$G(W_4)$ are digital 2-spheres.
Suppose that f: M→N is a homeomorphism of closed surfaces M and N. It is clear that if $W(M)=\{D_1,\ldots D_m\}$ is an LCL cover of M then $W(N)=\{f(D_1),\ldots f(D_m)\}$ is an LCL cover of N, and the intersection graphs G(W(M)) and G(W(N)) are isomorphic.

In medical imaging, regions of interest are often represented by 2-dimensional surfaces. Consider segmentation techniques, which enable to choose the shape and the size of individual elements within the framework of the proposed scheme in order to represent objects with fine anatomical details, and integrate geometric and topological features of closed surfaces into their digital models. For this purpose, we use an LCL segmentation of a closed surface.

**Definition 4.2**
Let M be a closed surface, an LCL collection $W(M)=\{D_1,D_2,\ldots D_s\}$ of 2-cells be a cover of M, and $G(W(M))$ be the intersection graph of $W(M)$. $G(W(M))$ is called the digital model of M in regard to $W(M)$.

**Theorem 4.3**
Let M be a closed surface, and an LCL collection $W(M)=\{D_0,D_1,\ldots D_k\}$ of 2-cells be a cover of $M=D_0\cup\ldots D_k$. Then the intersection graph $G(W(M))$ of $W(M)$ is a digital 2-manifold.

**Proof**
Suppose that $W(M)=\{D_0,D_1,\ldots\}$ is an LCL cover of M and consider the collection $U=\{D_1,D_2\ldots D_s\}$ containing all 2-cells, which intersect $D_0$, $D_0\cap D_i\neq\varnothing$, i=1,…s. Then the collection $V(\partial D_0)=\{C_1,C_2,\ldots C_s\}$, where $C_i=D_0\cap D_i$, i=1,…s, is an LCL collection of 1-cells according to proposition 2.3, and a cover of the circle $\partial D_0$. Therefore, the intersection graph $G(V(\partial D_0))$ is a digital 1-sphere according to theorem 4.1. Since collections U and $V(\partial D_0)$ are isomorphic according to proposition 2.4, then the intersection graphs $G(U)$ and $G(V(\partial D_0))$ of U and $V(\partial D_0)$ are isomorphic, i.e., $G(U)$ is a digital 1-sphere. Since $G(U)$ is the rim $O(x_0)$ of point $x_0$ in $G(W(M))$, then $O(x_0)$ is a digital 1-sphere. Similarly, $O(x_i)$ is a digital 1-sphere for any point $x_i\in G(W(M))$. Therefore, $G(W(M))=\{x_0,x_1,\ldots\}$ is a digital 2-manifols. □

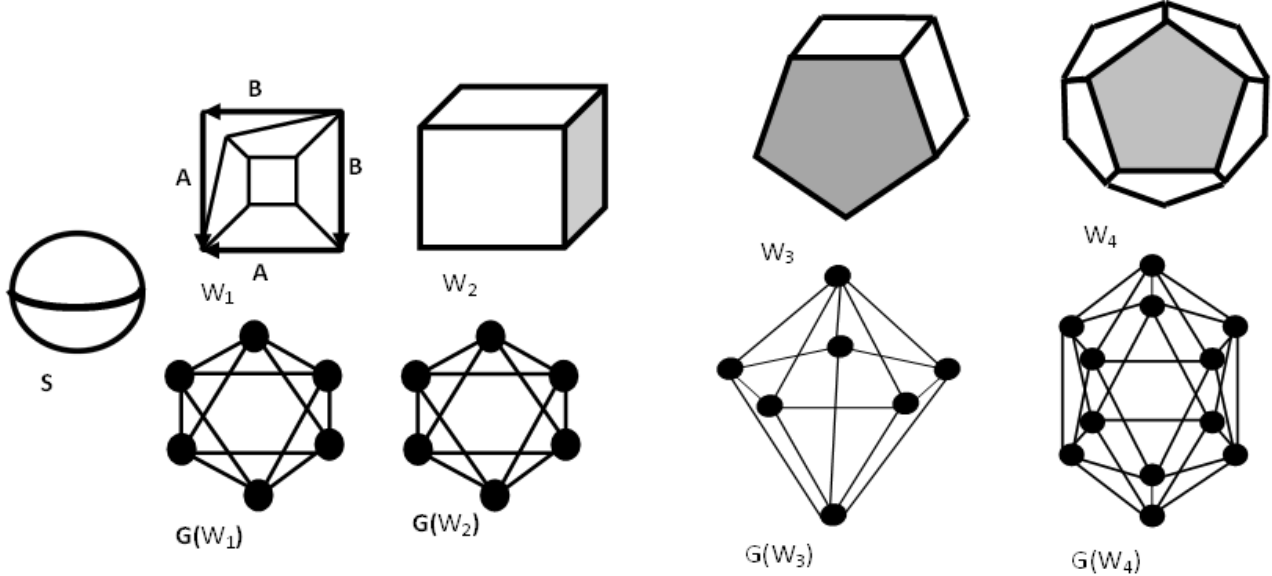

Figure 9. $W_1$ $W_2$ $W_3$ and $W_4$ are LCL covers of a continuous 2-sphere S. The Intersection graphs $G(W_1)$, $G(W_2)$, $G(W_3)$ and $G(W_4)$ are digital 2-spheres.

In fig. 10, $W_1$ is an LCL cover of a projective plane, $G(W_1)$ and $H(W_1)$ is the intersection graphs of $W_1$, $W_2$ is an LCL cover of a torus, $G(W_2)$ is the intersection graph of $W_2$. $G(W_1)$, $H(W_1)$ and $G(W_2)$ are digital 2-manifolds.
The following assertion is a consequence of proposition 3.3, theorem 4.2 and theorem 4.3.

**Theorem 4.4**
Let M be a closed surface, and collections $W(M)=\{D_1,D_2,\ldots D_m\}$ and $W(N)=\{E_1,E_2,\ldots E_n\}$ be LCL covers of M by 2-cells. Then:
- Digital models $G(W(M))$ and $G(W(N))$ of M are homotopy equivalent digital 2-manifolds.
- There is a compressed digital 2-manifold $C(M)$, which is a digital model of M, and which can be

obtained from G(W(M)) or G(W(N)) by sequential contracting simple pairs.

## 5. Classification of closed surfaces by digital tools

**Definition 5.1**
Let M be a closed surface, $W(M)=\{D_1,D_2,\ldots D_m\}$ be LCL covers of M by 2-cells, and $G(W(M))$ be the digital model of M in regard to W(M). Let C(M) be a compressed digital 2-manifold homotopy equivalent to G(W(M)). We say that C(M) is the compressed digital model of M, and the number of points of C(M) is the digital weight of M. Denote the digital weight of M by dw(M).

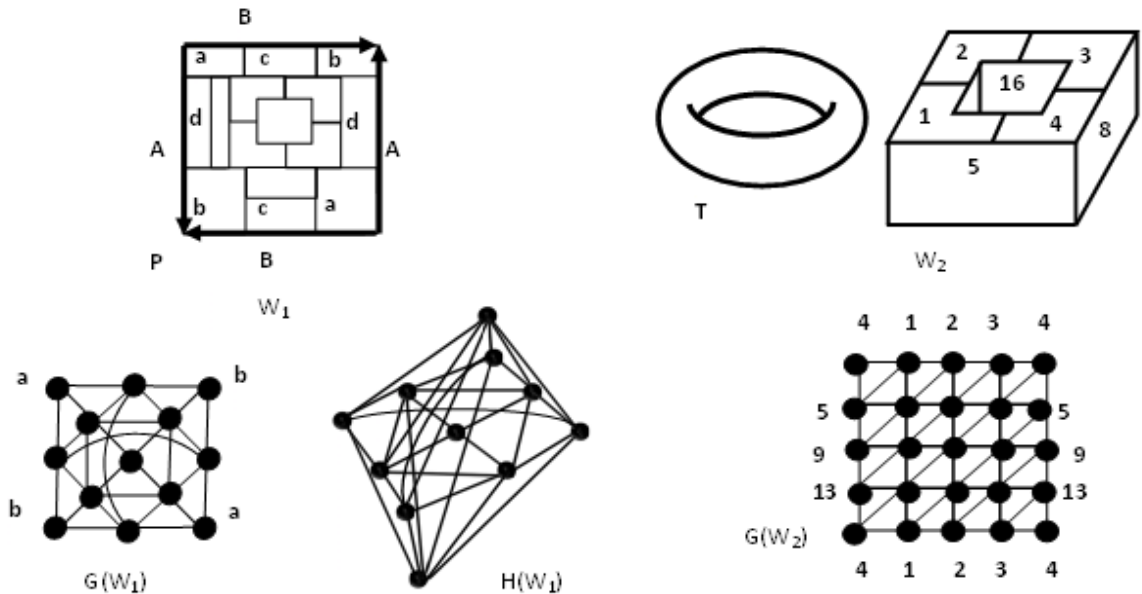

Figure 10. $W_1$ is an LCL cover of a projective plane P, intersection graphs $G(W_1)$ and $H(W_1)$ are isomorphic digital 2-projective planes. $W_2$ is an LCL cover of a torus T, the intersection graph $G(W_2)$ is a digital 2-torus.

Digital weight dw(M) and compressed digital model C(M) can be regarded as a topological invariants of a closed surface M in the following sense: if M is homeomorphic to N, then dw(M)=dw(N), and C(M) is homeomorphic to C(N). Evidently, C(M) and dw(M) can be used for a new classification of closed surfaces by digital tools. For a closed surface M, there is an LCL cover W(M) of M. Therefore, one can build the intersection graph G(W(M)), which is a digital 2-manifols, and which is the digital model of M. G(W(M)) can be converted to the compressed digital 2-manifols C(M), which is also is a digital model of M, and which is homotopy equivalent to G(W(M)). As it follows from the obtained results, dw(M) is the minimal number of elements contained in the LCL cover of M, or the minimal number of points necessary for building a digital model of M. Roughly speaking, the smaller dw(M) the simpler M.
As one can see from fig. 7, if M is a circuit, then dw(M)=4. In fig. 9-10, the minimal LCL covers of a sphere S, a projective plane P and a torus T, and compressed digital models G(S), G(P) and G(T) are shown. In table 2, digital weights of three closed surfaces are shown.

| Closed surface | Sphere S | Projective plane P | Torus T |
|---|---|---|---|
| Digital weight dw | 6 | 11 | 16 |

Table 1. Digital weights of closed surfaces

For the digital weight, we have dw(S)=6, dw(P)=11 and dw(T)=16. Thus, a sphere S is the simplest closed surface, because its digital model G(S) contains six points at least. The next closed surface by the digital weight is a projective plane P, for which G(P) contains eleven points at least. For a torus T, its digital model G(T) contains sixteen points at least. It turn out that the projective plane is simpler than the torus. In connection with the results, the following questions may arise: What are properties of the sequence of digital weights? For a given positive integer N, how many non-homeomorphic closed surfaces have the digital weight equal to N, or smaller than N?

Now we can describe a method for constructing digital models of closed surfaces and calculating their digital weights. Let M be a closed surface. Then:

- Build an LCL cover W(M) of M.
- Construct the intersection graph G(W(M)) of W(M).
- Convert G(W(M)) to the compressed form C(M) by sequential contracting simple pairs.
- Find the digital weight dw(M) of M, i.e., the number of points in C(M).

It is clear, that realizing this method is a computational problem, which is understood to be a task that is in principle capable to being solved by a computer.

## 6. Conclusion

This paper proposes a novel method for constructing digital models of continuous closed surfaces.

- We define an LCL cover of a closed surface M, and show that the intersection graph of any LCL cover of a given closed surface M is necessarily a digital 2-surface preserving the geometry and topology of M.
- It is shown that a closed surface M can be characterized by the compressed digital 2-surface C(M), and the digital weight dw(M), which is the number of points of C(M).
- A digital weight dw(M) and a compressed digital model C(M) can be regarded as a topological invariants of a closed surface. C(M) and dw(M) can be used for a new classification of closed surfaces by digital tools.
- We show that if closed surfaces M and N are homeomorphic, then dw(M)=dw(N), and their compressed digital models G(M) and G(N) are homeomorphic.